\newcommand{\changed}[1]{#1}

\documentclass[twocolumn,amsthm,secthm]{autart}    

\usepackage{graphicx}          

\usepackage{subfig}
\usepackage{amsmath}
\usepackage{amsfonts}
\usepackage{amssymb}  
\usepackage{bm}
\usepackage{siunitx}
\usepackage{xspace}
\usepackage{xfrac}
\usepackage{empheq}
\usepackage{enumerate}
\usepackage{xcolor}
\usepackage[round]{natbib}

\DeclareMathOperator{\sign}{sign}

\renewcommand{\epsilon}{\varepsilon}
\renewcommand{\phi}{\varphi}

\newcommand{\abs}[1]{\left\lvert #1 \right\rvert}
\newcommand{\norm}[2][]{\left\lVert #2 \right\rVert_{#1}}

\newcommand{\TT}{^{\mathrm{T}}}

\def\A{\mathbf{A}}

\def\F{\mathbf{F}}

\def\M{\mathbf{M}}

\def\c{\mathbf{c}}

\def\f{\mathbf{f}}
\def\g{\mathbf{g}}
\def\h{\mathbf{h}}

\def\x{\mathbf{x}}
\def\y{\mathbf{y}}
\def\z{\mathbf{z}}

\newcommand{\RR}{\mathbb{R}}

\newcommand{\AC}{AC\xspace}
\newcommand{\ACs}{AC$_*$\xspace}
\newcommand{\ACGs}{ACG$_*$\xspace}
\newcommand{\NN}{\mathbb{N}}

\renewcommand{\F}{\mathcal{F}}
\newcommand{\Sc}{\mathcal{S}}
\newcommand{\Ic}{\mathcal{I}}
\newcommand{\Jc}{\mathcal{J}}
\newcommand{\Hc}{\mathcal{H}}
\newcommand{\Dc}{\mathcal{D}}
\newcommand{\Bc}{\mathcal{B}}
\newcommand{\Ec}{\mathcal{G}}

\newcommand{\mex}{\hfill$\triangle$}

\newcommand{\pderiv}[2]{\frac{\partial{#1}}{\partial{#2}}}

\begin{document}

\begin{frontmatter}

\title{Generalized Filippov solutions for systems with prescribed-time convergence}

\thanks{The financial support by the Christian Doppler Research Association, the Austrian Federal Ministry for Digital and Economic Affairs and the National Foundation for Research, Technology and Development is gratefully acknowledged.
}

\author[IRTCD]{Richard Seeber}\ead{richard.seeber@tugraz.at}

\address[IRTCD]{Christian Doppler Laboratory for Model Based Control of Complex Test Bed Systems, Institute of Automation and Control, Graz University of Technology, Graz, Austria}

\begin{keyword}                           Filippov solution; prescribed-time system; unbounded right-hand side; time-varying systems
\end{keyword}

\begin{abstract}                          Dynamical systems with prescribed-time convergence \changed{sometimes} feature a right-hand side exhibiting a singularity at the prescribed convergence time instant.
In an open neighborhood of this singularity, classical absolutely continuous Filippov solutions may fail to exist, preventing indefinite continuation of such solutions.
This note introduces a generalized Filippov solution definition based on the notion of \changed{generalized} absolute continuity in the \changed{restricted} sense.
Conditions for the continuability of such generalized solutions are presented and it is shown, in particular, that generalized Filippov solutions of systems with an equilibrium that is attractive, in prescribed time or otherwise, can always be continued indefinitely.
\changed{The results are demonstrated by applying them to a prescribed-time controller for a perturbed second-order integrator chain.}
 \end{abstract}

\end{frontmatter}

\begingroup
\def\QED{$\blacksquare$}
\section{Introduction}

Regulation of a dynamical system's state to the origin is an important topic in control theory.
In that regard, the type of convergence of the state is of particular relevance.
While linear control achieves exponential convergence, more advanced approaches, such as homogeneous or sliding-mode controllers or observers, can achieve convergence in finite time \citep{roxin1966finite,levant2005homogeneity} or fixed time \citep{polyakov2011nonlinear}.
More recently, \cite{song2017time,holloway2019prescribed} proposed approaches achieving so-called prescribed-time convergence using time-varying gains that tend to infinity at the prescribed convergence time instant.
\changed{While not all prescribed-time approaches use such unbounded gains, see e.g. \citep{zhou2022fixed},} several other prescribed-time approaches based on \changed{possibly unbounded} time-varying gains have since been proposed, e.g., by \cite{espitia2021predictor,zhou2021prescribed,aldana2022redesign}.

In \cite{song2017time} and many other works, trajectories are only considered until the convergence time instant $T$, i.e., on a time interval of the form $[t_0,T)$.
In some scenarios, continuing trajectories beyond the convergence time instant is desirable, for example by switching to a sliding mode approach after the convergence time instant.
From a theoretical point of view, this raises the issue of continuability of solutions as defined by \cite{filippov1988differential}.

The present communique shows that Filippov solutions of systems featuring prescribed-time convergence may indeed fail to exist beyond the convergence time instant.
A generalized Filippov solution definition is then suggested that eliminates this issue.
In particular, it is shown that, with the generalized solution definition, trajectories of prescribed-time systems may always be continued indefinitely, i.e., to the time interval $[t_0, \infty)$, paving the way for theoretically sound combinations of prescribed-time systems with other control approaches.

After introducing Filippov solutions in Section~\ref{sec:filippov}, Section~\ref{sec:motivation} motivates the discussion with two illustrative examples.
Section~\ref{sec:generalized} then defines generalized Filippov solutions and discusses their properties.
Main Theorems~\ref{th:cont1} and \ref{th:cont2} in Section~\ref{sec:continuability} show formal results on the continuability of such solutions, \changed{and Section~\ref{sec:example} applies them in a control design example.}
Section~\ref{sec:conclusion} draws conclusions.

\textbf{Notation:}
$\RR$, $\RR_{\ge 0}$, $\RR_{>0}$, and $\NN$ denote reals, nonnegative reals, positive reals, and positive integers.
Convex closure and Lebesgue measure of a set $\mathcal{S} \subseteq \RR^n$ are denoted by $\overline{\operatorname{co}} \, \Sc$ and $\mu(\Sc)$.
An interval is any connected set $\Ic \subseteq \RR$ with $\mu(\Ic) \ne 0$.
For a function $f$ with domain $\Ic$ and $\Jc  \subseteq \Ic$, $f|_{\Jc}$ is the restriction of $f$ to the domain $\Jc$.
For $y \in \RR$, $\lfloor y \rfloor$ and $\lceil y \rceil$ are the largest and smallest integer, respectively, not greater or not smaller than $y$.
\changed{The set of all subsets of a set $\mathcal{S}$ is denoted by $2^{\mathcal{S}}$.}

\section{Filippov solutions}
\label{sec:filippov}

This communique studies definitions and existence of solutions of differential equations of the form
\begin{equation}
    \label{eq:sys}
    \dot \x = \f(\x, t)
\end{equation}
where $\f : \RR^n \times \RR_{\ge 0} \to \RR^n$ may be discontinuous in $\x$ and, additionally, exhibit singularities in $t$.
Systems of this form commonly occur in the context of prescribed-time systems, cf. e.g. \cite{song2017time,holloway2019prescribed,zhou2021prescribed}, specifically in the form of time-varying gains that exhibit a singularity at a prescribed time instant $T \in \RR_{> 0}$.

Solutions of \eqref{eq:sys} may be studied in terms of the associated differential inclusion
\vspace{-0.5\baselineskip}
\begin{equation}
    \label{eq:sys:inc}
    \dot \x \in \F(\x, t)
\vspace{-0.5\baselineskip}\end{equation}
\changed{with $\F : \RR^n \times \RR_{\ge 0} \to 2^{\RR^n}$}
obtained by applying the Filippov procedure, cf. \citep{filippov1988differential}, to the right-hand side of \eqref{eq:sys};
specifically,
\begin{equation}
    \label{eq:filippov}
\F(\x, t) = \bigcap_{\delta \in \RR_{> 0}} \bigcap_{\substack{\Sc \subset \RR^n \\ \mu(\Sc) = 0}} \overline{\operatorname{co}} \, \{ \f(\tilde\x, t) : \tilde \x \in \Bc_{\delta}(\x) \setminus \Sc \}
\end{equation}
with $\Bc_\delta(\x) = \{ \tilde \x \in \RR^n : \norm{\x - \tilde \x} < \delta \}$.
The definition of Filippov solutions of \eqref{eq:sys} is based on the notion of (local) absolute continuity.
\begin{defn}
    \label{def:ac}
    A function $\g : \RR \to \RR^n$ is called
    \begin{enumerate}[a)]
        \item 
            \emph{absolutely continuous} (AC) on a closed interval $\Ic$, if for each $\epsilon > 0$ there exists $\delta > 0$ such that $\sum_{i=1}^{m} \norm{\g(b_i) - \g(a_i)} < \epsilon$ holds for each finite collection of non-overlapping intervals $[a_i, b_i] \subseteq \Ic$, $i=1, \ldots, m$, that satisfies $\sum_{i=1}^{m} |b_i - a_i| < \delta$;
        \item
            \emph{locally absolutely continuous} (LAC) on an interval $\Ic$, if it is absolutely continuous on each compact interval $\Jc \subseteq \Ic$.
    \end{enumerate}
\end{defn}
\changed{It is also worth to recall that a function is absolutely continuous if and only if it is continuous, has bounded variation, and satisfies Lusin's N-property \citep{gordon1994integrals}.}
According to \cite{filippov1988differential}, solutions of \eqref{eq:sys}---which are called classical Filippov solutions hereafter---are typically defined as follows.
\begin{defn}
    Let $\Ic \subseteq \RR_{\ge 0}$ be an interval.
    A function $\x : \Ic \to \RR^n$ is said to be a \emph{classical solution} of \eqref{eq:sys:inc} or a \emph{classical Filippov solution} of \eqref{eq:sys}, if it is LAC on $\Ic$ and satisfies \eqref{eq:sys:inc} almost everywhere on $\Ic$.
\end{defn}

\section{Motivating examples}
\label{sec:motivation}

To motivate the discussion, consider the error dynamics of a prescribed-time differentiator given by \eqref{eq:sys} with 
\begin{equation}
    \label{eq:exmp:f}
    \f(\x, t) = \begin{bmatrix}
        -h_1(x_1, t) + x_2 \\
        -h_2(x_1, t)
    \end{bmatrix},
\end{equation}
correction functions \changed{$h_1, h_2 : \RR \times \RR_{\ge 0} \to \RR$ defined as}
\begin{subequations}
    \label{eq:exmp:g}
    \begin{align}
        h_1(\changed{x_1}, t) &= \begin{cases}
            \frac{4}{T-t} \changed{x_1} & t \in [0,T) \\
            k_1 \sqrt{\abs{\changed{x_1}}} \sign(\changed{x_1}) & t \in [T,\infty)
        \end{cases} \\
        h_2(\changed{x_1}, t) &= \begin{cases}
            \frac{2}{(T-t)^2}\changed{x_1}  + \frac{1}{(T-t)^4} \changed{x_1} & t \in [0,T) \\
            k_2 \sign(\changed{x_1}) & t \in [T, \infty),
        \end{cases}
    \end{align}
\end{subequations}
and positive constants $k_1, k_2, T \in \RR_{>0}$.
On the time interval $[0,T)$, these error dynamics correspond to a linear time-varying differentiator of similar structure\footnote{Note that the example is \emph{not} a special case of the observers in \citep{holloway2019prescribed}, but nonetheless corresponds to an observer with prescribed-time convergence.} as the observers proposed, e.g., by \cite{holloway2019prescribed}, which at $t = T$ is switched to the robust exact super-twisting differentiator due to \cite{levant1998robust}.
The next example shows that, beyond the formal similarity to the systems in \citep{holloway2019prescribed}, this system indeed features prescribed-time convergence, but classical Filippov solutions fail to exist in any (two-sided) neighborhood of the prescribed convergence time instant $T$, where $\f$ (and thus also $\F$) exhibits a singularity.

\begin{exmp}
    \label{exmp:bv}
    Consider system \eqref{eq:sys} with previously defined right-hand side \eqref{eq:exmp:f}--\eqref{eq:exmp:g}.
    Note that $\f(\cdot, t)$ is globally Lipschitz continuous on every compact subset of $[0,T)$.
    Hence, for every $\x_0 \in \RR^2$ and $t_0 \in [0,T)$, the function $\x : [t_0,T) \to \RR^2$ given by $\x(t) = \M(T-t) \M(T-t_0)^{-1} \x_0$ with
    \begin{equation}
        \label{eq:M:bv}
        \M(\tau) = \begin{bmatrix}
            \tau^3 \sin \frac{1}{\tau} & \tau^3 \cos \frac{1}{\tau} \\
            \tau^2 \sin \frac{1}{\tau} + \tau \cos \frac{1}{\tau} & \tau^2 \cos \frac{1}{\tau} - \tau \sin \frac{1}{\tau}
        \end{bmatrix}
    \end{equation}
    is the unique classical solution with $\x(t_0) = \x_0$; \changed{for $T = 1$,} it is  depicted in Fig.~\ref{fig:bv}.
    However, its only continuous extension to $[t_0,T]$ with $\x(T) = \bm{0}$ is not absolutely continuous, because it does not have bounded variation.
    \changed{Indeed, the total variation of the term $g(\tau) = \tau \cos \frac{1}{\tau}$ with $\tau = T-t \in (0,T-t_0]$, for example, is bounded from below by
        \begin{align}
            \hspace{-0.55em}
\sum_{i=I}^{\infty} \left|g\left(\frac{1}{i \pi}\right) - g\left(\frac{1}{(i+1) \pi}\right)\right| &= \sum_{i=I}^{\infty} \left|\frac{(-1)^i}{i \pi} - \frac{(-1)^{i+1}}{(i+1)\pi}\right| \nonumber \\
            &\ge \sum_{i=I}^{\infty} \frac{2}{(i+1)\pi} = \infty,
        \end{align}}\changed{wherein $I = \lceil\frac{1}{(T - t_0) \pi} \rceil$.}
    Hence, every (nontrivial) continuous function defined on $(T-\epsilon, T+\epsilon)$ with $\epsilon > 0$ that satisfies the corresponding inclusion \eqref{eq:sys:inc} almost everywhere is \emph{not locally absolutely continuous}.
As a consequence, no nontrivial classical solutions exist on intervals of this form, i.e., classical solutions cannot be continued beyond the singularity occuring at $T$.\mex
\end{exmp}
\begin{figure}[tbp]
    \centering
    \includegraphics{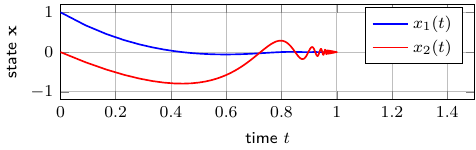}
    \caption{\changed{Classical Filippov} solution from Example~\ref{exmp:bv} with \changed{$T = 1$ and initial conditions} $t_0 = 0$, $\changed{\x(t_0) = } \x_0 = [1 \quad 0]\TT$.}
    \label{fig:bv}
\end{figure}

The next example shows that, for systems that do \emph{not} feature prescribed-time convergence, continuability of bounded solutions can fail due to yet another problem.
\begin{exmp}
    \label{exmp:cont}
    Consider system \eqref{eq:sys} with
    \begin{equation}
        \f(\x,t) = \begin{cases}
            \A(T-t) \x & t \in [0, T) \\
            \A(T) \x & t \in [T, \infty),
        \end{cases} \,
        \A(\tau) = 
        \begin{bmatrix}
            -\frac{1}{\tau} & 1 \\
            -\frac{1}{\tau^2} & 0
        \end{bmatrix},
    \end{equation}
    with $T \in \RR_{> 0}$.
For $\x_0 \in \RR^2$, $t_0 \in [0,T)$, the unique (and bounded) classical solution is given by $\x : [t_0, T) \to \RR^2$ defined as $\x(t) = \M(T-t) \M(T-t_0)^{-1} \x_0$ with
    \begin{equation}
        \M(\tau) = \begin{bmatrix}
            \tau \sin \ln \tau & \tau \cos \ln \tau \\
            - \cos \ln \tau & \sin \ln \tau
        \end{bmatrix}.
    \end{equation}
    Fig.~\ref{fig:cont} \changed{exemplarily depicts it for $T = 1$}.
    Since $\lim_{t \to T} \x(t)$ does not exist for $\x_0 \ne \bm{0}$, every nontrivial function that satisfies \eqref{eq:sys} almost everywhere on $(T-\epsilon, T+\epsilon)$ with $\epsilon > 0$ is \emph{discontinuous} at $t = T$.
    As a consequence, no nontrivial classical solutions exist on intervals of this form.\mex
\end{exmp}
\begin{figure}[tbp]
    \centering
    \includegraphics{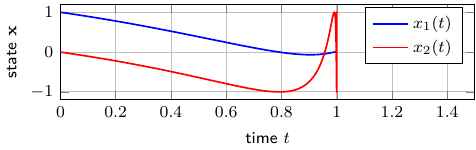}
    \caption{\changed{Classical Filippov} solution from Example~\ref{exmp:cont} with \changed{$T = 1$ and initial conditions} $t_0 = 0$, $\changed{\x(t_0) =} \x_0 = [1 \quad 0]\TT$.}
    \label{fig:cont}
\end{figure}

\changed{Recall again that absolute continuity is equivalent to the three properties continuity, boundedness of variation, and Lusin's N-property.}
While Lusin's property cannot conceivably be lost, solutions near $t = T$ fail to exist due to a lack of bounded variation in Example~\ref{exmp:bv} and due to a lack of continuity in Example~\ref{exmp:cont}.
It is clear that, in the latter example, any attempt to continuously continue solutions must fail.
However, intuitively, solutions in the former example should be continuable by the zero solution.

The present communique addresses this issue by proposing a generalized Filippov solution definition, which guarantees existence and continuability of solutions under very mild conditions.
In particular, it will be shown that the proposed generalized Filippov solutions are always continuable to \changed{the interval} $[t_0, \infty)$ for systems whose origin is attractive, in prescribed time or otherwise.

\section{Generalized Filippov solutions}
\label{sec:generalized}

In the following, a generalized solution definition for Filippov type systems is introduced based on the class of functions that are generalized absolutely continuous in the restricted sense (\ACGs) as originally introduced by \cite{lusin1912proprietes}, cf. \citep{saks1937theory}.
\begin{defn}[{\cite[Definition 6.1]{gordon1994integrals}}]
    \label{def:acg}
    A function $\g : \RR \to \RR^n$ is said to be
    \begin{enumerate}[a)]
        \item 
            \emph{absolutely continuous in the restricted sense} (\ACs) on a set $\Sc \subseteq \RR$, if for each $\epsilon > 0$ there exists $\delta > 0$ such that $\sum_{i=1}^{m} \sup_{a,b \in [a_i,b_i]} \norm{\g(b) - \g(a)} < \epsilon$ for each finite collection of non-overlapping intervals $[a_i, b_i]$, $i=1, \ldots, m$, that satisfies $a_i, b_i \in \Sc$ and $\sum_{i=1}^{m} |b_i - a_i| < \delta$;
        \item
            \emph{generalized absolutely continuous in the restricted sense} (\ACGs) on a set $\Sc \subseteq \RR$, if \changed{$\g|_{\Sc}$} is continuous on $\Sc$ and $\Sc$ can be written as the union of countably many sets on each of which $\g$ is \ACs.
    \end{enumerate}
\end{defn}

\subsection{Definition of generalized solutions}

Like LAC functions, also \ACGs functions are differentiable almost everywhere, cf. \cite[Corollary 6.19]{gordon1994integrals}.
This leads to the following definition.

\begin{defn}
    Let $\Ic \subseteq \RR_{\ge 0}$ be an interval.
    A function $\x : \Ic \to \RR^n$ is said to be a \emph{generalized solution} of \eqref{eq:sys:inc} or a \emph{generalized Filippov solution} of \eqref{eq:sys}, if it is \ACGs \changed{on $\Ic$} and satisfies \eqref{eq:sys:inc} almost everywhere \changed{on $\Ic$}.
\end{defn}
\changed{\begin{rem}
    The proposed definition may also be extended to different solution definitions of discontinuous systems, such as Utkin solutions or Aizerman-Pyatnitskii solutions, cf. \citep{polyakov2014stability} for an overview, by analogously relaxing the absolute continuity requirement in those definitions to generalized absolute continuity in the restricted sense.
\end{rem}
}
It is worth noting that, while \changed{LAC} functions are associated to the Lebesgue integral, \ACGs functions are associated to the Denjoy integral\footnote{Or to the equivalent Perron or Henstock-Kurzweil integral.}.
Specifically, a function $\h : \Ic \to \RR^n$ is Denjoy integrable iff there exists an \ACGs function $\g : \Ic \to \RR^n$ such that $\dot \g(t) = \h(t)$ almost everywhere \changed{on $\Ic$}, cf. e.g. \cite[Definition 7.1]{gordon1994integrals}.

A similar solution definition for \emph{ordinary differential equations}, using the Perron integral, is considered by \cite{kurzweil1990ordinary}, who also study some general criteria for the existence of such solutions, cf. \cite{schwabik1990generalized} and references therein.
Here, the considered generalized solution definition for Filippov systems is studied in the context of prescribed-time systems, the main result being conditions for continuability of generalized solutions for all systems of this type.

\subsection{Properties of generalized solutions}

The following lemma, proven in the appendix, shows some connections between the different classes of functions \changed{\cite[cf. also][]{gordon1994integrals}}.
\begin{lem}
    \label{lem:acg}
    Consider an interval $\Ic$, a \changed{continuous} function $\g : \changed{\Ic} \to \RR^n$, and a countable collection of intervals $(\Ic_j)$, $j \in \NN$, with $\Ic = \cup_{j=1}^{\infty} \Ic_j$.
    Then, the following statements are true\footnote{Note that item (\ref{it:acg}) needs the Axiom of Choice, unless the collection $(\Ic_j)$ is finite.}:
\begin{enumerate}[a)]
        \item
            \label{it:ac}
            for closed $\Ic$, $\g$ is AC on $\Ic$ if and only if it is \ACs on $\Ic$;
\item
            \label{it:lac}
            if $\g$ is LAC on $\Ic$, then it is \ACGs on $\Ic$;
        \item
            \label{it:acs}
            if $\g$ is \ACs on $\Ic$, then it is LAC on $\Ic$;
        \item
            \label{it:acg}
            if $\g$ is \ACGs on each $\Ic_j$, then it is \ACGs on $\Ic$.
    \end{enumerate}
\end{lem}

From item (\ref{it:lac}), the following proposition is obvious.
\begin{prop}
    \label{prop:embed}
    Every classical solution of \eqref{eq:sys:inc} is also a generalized solution.
\end{prop}
The next proposition shows how classical solutions can be recovered from generalized solutions using item (\ref{it:acs}).
\begin{prop}
    \label{prop:decomp}
    Let $\x : \Ic \to \RR^n$ be a generalized solution of \eqref{eq:sys:inc}.
    Then, for every interval $\Jc \subseteq \Ic$, an open interval $\Hc \subseteq \Jc$ exists such that $\x|_\Hc$ is a classical solution.
\end{prop}
\begin{proof}
    Let $\tilde \Jc$ be a closed interval contained in $\Jc$.
    Since $\tilde \Jc$ is a perfect set, it contains an open subinterval $\Hc$ on which $\x$ is \ACs according to \cite[Theorem 6.1]{gordon1994integrals}.
    Hence, $\x|_\Hc$ is LAC according to Lemma~\ref{lem:acg}, item (\ref{it:acs}), and is thus a classical solution.
\end{proof}

Item~(\ref{it:acg}) shows that concatenating countably many generalized solutions again yields a generalized solution.
\begin{prop}
    \label{prop:concat}
    Let $(\Ic_j)$, $(\x_j)$, $j \in \NN$, be a countable collection of intervals and corresponding classical or generalized solutions $\x_j : \Ic_j \to \RR^n$ of \eqref{eq:sys:inc} with the property that $\Ic = \cup_{j=1}^{\infty} \Ic_j$ is an interval and that, for every pair of integers $i, j \in \NN$, \changed{$\lim_{t \to \tau} [\x_j(t) - \x_i(t)] = \bm{0}$ for all $\tau \in \overline{\operatorname{co}} \, \Ic_i \cap \overline{\operatorname{co}}\, \Ic_j$}.
    Then, $\x : \Ic \to \RR^n$ defined as $\x(\tau) = \x_j(\tau)$ for $\tau \in \Ic_j$ is a generalized solution of \eqref{eq:sys:inc}.
\end{prop}
\begin{proof}
    Clearly, $\x$ is well-defined and continuous by construction.
    Moreover, from Proposition~\ref{prop:embed}, each $\x_j$ is a generalized solution and hence \ACGs on $\Ic_j$.
Thus, $\x$ is \ACGs on $\Ic$ due to Lemma~\ref{lem:acg}, item (\ref{it:acg}).
    Since it also satisfies the inclusion almost everywhere on each $\Ic_j$, it is a generalized solution of \eqref{eq:sys:inc}.
\end{proof}

\section{Continuability of generalized solutions}
\label{sec:continuability}

Formal conditions for the continuability of generalized Filippov solutions are now shown.
First, general systems with a singularity at some time instant $T \in \RR_{> 0}$ are considered.
Then, prescribed-time systems are addressed, by showing indefinite continuability of generalized solutions for systems with an attractive equilibrium.

\subsection{General systems with singularities}

\begin{defn}
    \label{def:interval}
    A solution $\x : \Ic \to \RR^n$ is said to be \emph{maximal}, if there exists no solution $\bar \x : \bar \Ic \to \RR^n$ with $\bar \Ic \cap \Ic \ne \{ \}$ and $\sup \bar \Ic > \sup \Ic$ satisfying $\x(t) = \bar \x(t)$ for all $t \in \Ic \cap \bar \Ic$.
\end{defn}

The next theorem shows that generalized solutions that are continuous at a singularity at some time instant $T$ can be continued beyond $T$ subject to the usual Filippov conditions, cf. \cite{filippov1988differential}.
\begin{thm}
    \label{th:cont1}
    Let $T,\epsilon \in \RR_{> 0}$ and
consider the inclusion \eqref{eq:sys:inc}.
    Suppose that $\F : \RR^n \times \RR_{\ge 0} \to 2^{\RR^n}$ is upper semicontinuous on every compact subset of $\Ec = \RR^n \times [T,T+\epsilon)$ and that the set $\F(\x,t)$ is nonempty, compact, and convex for all $(\x,t) \in \Ec$.
    Let $\x : \Ic \to \RR^n$ be a maximal generalized solution of \eqref{eq:sys:inc}.
    If $\sup \Ic \ge T$ and $\lim_{t \to T} \x(t)$ exists, then there exists $\delta \in \RR_{> 0}$ such that $\sup \Ic \ge T+\delta$.
\end{thm}
\begin{rem}
    Note that the theorem imposes conditions on $\F$ only \emph{after} the singularity, and requires existence of the generalized solution only \emph{before} and its continuity \emph{at} the singularity (cf., however, Example~\ref{exmp:cont}).
\end{rem}
\begin{proof}
    Suppose to the contrary that $\sup \Ic = T$, and let $\bar \x_0 = \lim_{t \to T} \x(t)$.
    According to \cite[Theorem 7.1]{filippov1988differential}, there exists a classical solution $\bar \x : [T, T+\delta] \to \RR^n$ with $\bar \x(T) = \bar \x_0$ and some $\delta \in \RR_{> 0}$.
    Due to Proposition~\ref{prop:concat}, there then exists a generalized solution defined on $\Ic \cup [T, T+\delta]$ which coincides with $\x$ on $\Ic$.
    This contradicts the fact that $\x$ is maximal.
\end{proof}

Theorem~\ref{th:cont1} already shows local continuability of solutions in Example~\ref{exmp:bv}.
For systems with attractive equilibria, a more powerful result, allowing to continue solutions indefinitely, is shown in the next section.

\subsection{Systems with attractive equilibria}

\begin{defn}
    \label{def:attractive}
    The origin of the inclusion \eqref{eq:sys:inc} is said to be an \emph{attractive equilibrium} in the classical (generalized) sense, if $\bm{0} \in \F(\bm{0}, t)$ holds for almost all $t \in \RR_{\ge 0}$, and all maximal classical (generalized) solutions $\x : \Ic \to \RR^n$ satisfy $\lim_{t \to \sup \Ic} \x(t) = \bm{0}$.
\end{defn}

To establish the connection to prescribed-time systems, it is first shown that such systems always have an attractive equilibrium if the zero solution is unique.
\begin{prop}
    \label{prop:equilibrium}
    Consider the inclusion \eqref{eq:sys:inc}.
    Suppose that \changed{for all $t_0 \in \RR_{\ge 0}$} the unique classical (generalized) solution \changed{$\bar \x$} with $\changed{\bar\x}(t_0) = \bm{0}$ satisfies $\changed{\bar \x}(t) = \bm{0}$ for all $t \ge t_0$, and that every maximal classical (generalized) solution $\x \changed{: \Ic \to \RR^n}$ fulfills $\lim_{t \to T} \x(t) = \bm{0}$ for some $T \in \RR_{\ge 0} \cup \{ \infty \}$.
    Then, the origin is an attractive equilibrium in the classical (generalized) sense.
\end{prop}
\begin{proof}
    \changed{Since the zero solution $\bar \x(t) = \bm{0}$ is a solution, the inclusion $\bm{0} \in \F(\bm{0}, t)$ holds for almost all $t \in \RR_{\ge 0}$.
        Consider now any maximal solution $\x : \Ic \to \RR^n$ satisfying $\lim_{t \to T} \x(t) = \bm{0}$ by assumption.
        If $T = \sup \Ic$, then there is nothing to prove.
        Otherwise, $T \in \Ic$ is finite and $\x(T) = \bm{0}$ holds by continuity of $\x$.
        Since the zero solution is unique, then $\x(t) = \bm{0}$ holds for all $t \in [T,\infty) \cap \Ic$, implying $\lim_{t \to \sup \Ic} \x(t) = \bm{0}$.
    }
\end{proof}

Denote by $\Dc$ the set of time instants where \eqref{eq:sys:inc} exhibits a singularity, i.e., where $\F$ is not locally bounded\footnote{\changed{Note that a set-valued function $\mathcal{F}$ is called locally bounded on a set $\mathcal{G}$, if for all $\y \in \mathcal{G}$ there exist constants $\varepsilon, M \in \RR_{> 0}$ such that $\sup_{\z \in \mathcal{F}(\tilde\y)} \norm{\z} \le M$ for all $\tilde \y \in \mathcal{G}$ with $\norm{\y - \tilde \y}< \varepsilon$.}}.
It will be shown that, if $\Dc$ does not have cluster points, then \emph{generalized} solutions of systems with attractive equilibria in the \emph{classical} sense are always continuable indefinitely.
If $\Dc$ is moreover finite, attractive equilibria in the classical and generalized sense will be shown to be \emph{equivalent}.
To that end, the following auxiliary lemma is used, which is proven in the appendix.
\begin{lem}
    \label{lem:extract}
    Let $\Dc \subset \RR_{\ge 0}$ and consider the inclusion \eqref{eq:sys:inc}.
    Suppose that $\Dc$ has no cluster points and that $\F$ is locally bounded on $\Ec = \RR^n \times (\RR_{\ge 0} \setminus \Dc)$.
    Let $\x : \Ic \to \RR^n$ be a maximal generalized solution.
    Then, if either $\sup \Ic$ or $\sup \Dc$ are finite, an interval $\Jc \subseteq \Ic$ with $\sup \Jc = \sup \Ic$ exists such that $\x|_{\Jc}$ is a maximal classical solution.
\end{lem}
\begin{thm}
    \label{th:cont2}
    Let $\Dc\subset \RR_{\ge 0}$ and inclusion \eqref{eq:sys:inc} satisfy the conditions of Lemma~\ref{lem:extract}, and suppose that the origin is an attractive equilibrium in the \emph{classical} sense.
    Then, every maximal \emph{generalized} solution $\x : \Ic \to \RR^n$ satisfies $\sup \Ic = \infty$.
    If, moreover, $\sup \Dc$ is finite, then the origin is an attractive equilibrium in the \emph{generalized} sense.
\end{thm}
\begin{proof}
    Let $\x : \Ic \to \RR^n$ be a maximal generalized solution and suppose to the contrary that $\sup \Ic = T$ is finite.
    From Lemma~\ref{lem:extract} there exists an interval $\Jc$ with $\sup \Jc = T$ such that $\x|_{\Jc}$ is a maximal classical solution; hence, $\lim_{t \to T} \x(t) = \bm{0}$.
    But then the generalized solution $\x$ can be continued with the zero solution using Proposition~\ref{prop:concat}, a contradiction.
    Hence, $\sup \Ic = \infty$.
    If, moreover, $\sup \Dc$ is finite, then from Lemma~\ref{lem:extract} there exists an interval $\Jc$ with $\sup \Jc = \infty$ such that $\x|_{\Jc}$ is a maximal classical solution.
    Hence, $\lim_{t \to \infty} \x(t) = \bm{0}$.
\end{proof}

\section{\changed{Design Example: Prescribed-Time Control}}
\label{sec:example}

\changed{In the following, a class of prescribed-time controllers for a second-order system is designed that utilizes the proposed solution definition.
Consider the perturbed double integrator $\dot x_1 = x_2, \dot x_2 = u + w$ with a perturbation $w$ bounded according to $|w(t)| \le W \in \RR_{\ge 0}$ for all $t \in \RR_{\ge 0}$ subject to the control law
    \begin{equation}
        \label{eq:exmp:u}
        u(\x, t) = \begin{cases}
            - \left[\frac{\gamma^2}{(T-t)^4} + \frac{6}{(T-t)^2}\right] x_1 - \frac{4}{T-t} x_2 & t < T \\
            - k_1 \sign(x_1) - k_2 \sign(x_2) & t \ge T
        \end{cases}
\end{equation}
with prescribed time $T \in \RR_{> 0}$ and positive parameters $\gamma, k_1, k_2 \in \RR_{> 0}$.

Without perturbation, i.e., for $w(t) \equiv 0$, one non-trivial Filippov solution of the closed loop on the time interval $[0,T)$ is, for example, given by
\begin{align}
    x_1(t) &= (T - t)^3 \sin \frac{\gamma}{T-t}, \\
    x_2(t) &= (T-t) \gamma \cos\frac{\gamma}{T-t} - 3 (T-t)^2 \sin\frac{\gamma}{T-t}. \nonumber
\end{align}
Similar to Example~\ref{exmp:bv}, boundedness of variation is lost near $T$, and hence this solution cannot cross that singularity as a classical Filippov solution.
Nevertheless, the following proposition and its proof show that the considered control law achieves prescribed-time convergence on $[0, T)$ and finite-time convergence on $[T, \infty)$, and thus renders the origin attractive in the \emph{classical} sense.
\begin{prop}
    Let $\gamma, T \in \RR_{> 0}$, $W \in \RR_{\ge 0}$ and suppose that $k_1 > k_2 + W > 2W$.
    Consider the closed loop $\dot \x = \f(\x, t)$ defined via $\dot x_1 = x_2, \dot x_2 = u(\x, t) + w(t)$ with $u$ defined in \eqref{eq:exmp:u} and arbitrary Lebesgue measurable disturbance $w : \RR_{\ge 0} \to [-W, W]$.
    Then, the closed loop's origin is an attractive equilibrium in the classical sense.
\end{prop}
\begin{rem}
Using this proposition and Theorem~\ref{th:cont2} with $\mathcal{D} = \{ T \}$, indefinite continuability of \emph{generalized} solutions and attractivity of the origin in the \emph{generalized} sense may be concluded for the perturbed closed loop.
\end{rem}
\vspace{-0.2\baselineskip}
\begin{proof}
    Let $t_0 \in \RR_{\ge 0}$ and consider maximal classical Filippov solutions $\x : \Ic \to \RR$ of the closed loop with $t_0 \in \Ic$.
    Distinguish the cases $t_0 \ge T$ and $t_0 \in [0, T)$.
    In the first case, $\sup \Ic = \infty$ and $\lim_{t \to \infty} \x(t) = \bm{0}$ follows from finite-time stability of the twisting algorithm, cf. \citep{levant2007principles}.
    In the second case, $[t_0, T) \subseteq \Ic$ and $\lim_{t \to T} \x(t) = \bm{0}$, i.e., convergence in prescribed time $T$, will be shown.
    Solution existence on each compact subinterval of $[t_0, T)$ follows from the fact that the closed loop is an affine time-varying system satisfying a global Lipschitz condition on such a subinterval.
    To show prescribed-time convergence, define the abbreviation $q(\x, t) = \frac{3 x_1}{(T-t)^2} + \frac{x_2}{T-t}$ and consider the function $Q : \RR^2 \times [0, T) \to \RR_{\ge 0}$ defined as
    \begin{equation}
        Q(\x, t) = \frac{\gamma^2 x_1^2}{(T-t)^6} + q(\x, t)^2,
\end{equation}
    whose time derivative $\dot Q = \pderiv{Q}{\x}\f+\pderiv{Q}{t}$ along closed-loop trajectories on the time interval $[0, T)$ is given by
$
        \dot Q(\x, t) = \frac{2 w(t) q(\x, t)}{T-t}.
$
    Define $V : \RR^2 \times [0, T) \to \RR_{\ge 0}$ as $V(\x, t) = (T-t) Q(\x, t)$, whose time-derivative satisfies
    \begin{align}
        \dot V(\x, t) &= - Q(\x,t) + 2 w(t) q(\x, t) \nonumber \\
        &\le - Q(\x, t) + 2 w(t) q(\x, t) + \left[2 w(t) - \frac{q(\x, t)}{2}\right]^2 \nonumber \\
&\le - \frac{3}{4} Q(\x, t) + 4 w(t)^2 \le - \frac{3}{4} \frac{V(\x, t)}{T-t} + 4 W^2 \nonumber \\
        &\le - \frac{3}{4} \frac{V(\x, t)}{T} + 4 W^2.
\end{align}
    As a consequence, $V$ is uniformly bounded on $[t_0, T)$; specifically,
$
        V(\x(t), t) \le \max\{ V(\x(t_0), t_0), \frac{4 W}{\sqrt{3 T}} \}
$
    holds for all $t \in [t_0, T)$.
The claim $\lim_{t \to \infty} \x(t) = \bm{0}$ is now obtained by contradiction as follows.
    Suppose that a sequence $(t_k)$, $t_k \in [t_0, T)$ and a nonzero vector $\c \in \RR^2$ exist with $\lim_{k\to \infty} t_k = T$ and $\lim_{k \to \infty} \x(t_k) = \c$.
    Since $\lim_{t \to T} V(\c, t) = \infty$ for each fixed, nonzero $\c$,
    \begin{align}
        \limsup_{t\to\infty} V(\x(t), t) &\ge \lim_{k \to \infty} V(\x(t_k), t_k) \nonumber \\
        &= \lim_{k \to \infty} V(\c, t_k) = \infty
    \end{align}
    then follows, which contradicts the fact that $V(\x(t), t)$ is uniformly bounded on $[t_0, T)$.
\end{proof}
\begin{rem}
It is worth noting that the arguments of the previous proof may also be used to show (non-uniform) Lyapunov stability, provided that generalized Filippov solutions are considered.
Arguing this in formal detail is beyond the scope of the present paper, however.
\end{rem}
}

\section{Conclusion}
\label{sec:conclusion}

It was shown that Filippov solutions of prescribed-time systems may in some cases fail to be continuable beyond the prescribed convergence time instant.
To overcome this restriction, a definition of generalized Filippov solutions was introduced.
Indefinite continuability of such generalized solutions was shown for systems with an attractive equilibrium and thus, in particular, for all systems featuring prescribed-time convergence.
These results pave the way for theoretically sound combinations of new or existing prescribed-time approaches with other control techniques, such as sliding mode control.
\changed{This was demonstrated by designing a combined prescribed-time/sliding-mode controller for a perturbed second-order integrator chain.}

\appendix

\section{Proofs}
\begin{proof}[Proof of Lemma~\ref{lem:acg}]
For item (\ref{it:ac}),
note that if $\Sc$ is a closed interval in Definition~\ref{def:acg}, then $[a_i, b_i] \subseteq \Sc$ iff $a_i, b_i \in \Sc$.
The fact that \ACs implies AC then follows from $\norm{\g(b_i) - \g(a_i)} \le \sup_{a,b \in [a_i, b_i]} \norm{\g(b) - \g(a)}$.
Conversely, suppose that $f$ is AC,
let $\epsilon > 0$ and let $\delta > 0$ be as in Definition~\ref{def:ac}.
Consider any finite collection $\Ic_i = [a_i, b_i]$ of non-overlapping intervals as in Definition~\ref{def:acg}.
Since $f$ is continuous and $\Ic_i$ is compact, there exist $c_i, d_i \in \Ic_i \subseteq \Sc$ with $c_i < d_i$ such that $\sup_{a,b \in \Ic_i} \norm{\g(b) - \g(a)} = \norm{\g(d_i) - \g(c_i)}$.
Noting that $\sum_{i=1}^{m} |d_i - c_i| \le \sum_{i=1}^m |b_i - a_i| < \delta$ then yields $\sum_{i=1}^{m} \sup_{a,b \in \Ic_i} \norm{\g(b) - \g(a)} < \epsilon$ from Definition~\ref{def:ac}.
For item (\ref{it:lac}), since $\Ic$ is an interval, it can be written as the countable union of compact intervals $\Jc_j$, on each of which $\g$ is AC.
Since each $\Jc_j$ is closed, $\g$ being AC implies $\g$ being \ACs on $\Jc_j$ due to item (a).
Hence, $\g$ is \ACGs on $\Ic$ by definition.
Item (\ref{it:acs}) follows from the fact that $\g$ is \ACs on every subset of $\Ic$, and hence \AC on all compact subintervals of $\Ic$ due to item (\ref{it:ac}).
To see item (\ref{it:acg}), for each $j$, define $\Hc_j$ as the countable set of sets according to Definition~\ref{def:acg}, with union $\Ic_j = \cup_{\Sc \in \Hc_j} \Sc$, on which $\g$ is \ACs.
Then, the set of sets $\Hc = \cup_{j=1}^{\infty} \Hc_j$, being the countable union of countable sets, is countable, and $\g$ is \ACs on each $\Sc \in \Hc$.
Since $\cup_{\Sc \in \Hc} \Sc = \cup_{j=1}^{\infty} \Ic_j = \Ic$, $\g$ is \ACGs on $\Ic$ by definition.
\end{proof}
\begin{proof}[Proof of Lemma~\ref{lem:extract}]
    Let $T = \sup \Ic \in \RR_{\ge 0} \cup \{ \infty \}$ and choose $\theta \in \Ic$ such that $\Dc \cap (\theta, T) = \emptyset$.
    From Proposition~\ref{prop:decomp} there exists an open interval $\Hc \subseteq (\theta,T)$, such that $\x|_{\Hc}$ is a classical solution.
    Choose any $\tau_1 \in \Hc$ and the largest $\tau_2 \in (\tau_1, T]$ such that $\x$ is LAC on $\Jc = [\tau_1, \tau_2)$.
    To show $\sup \Jc = \sup \Ic$, which also implies that $\x|_{\Jc}$ is maximal because otherwise $\x$ could be extended using Proposition~\ref{prop:concat}, suppose to the contrary that $\tau_2 < T$.
    Consider the compact interval $\bar \Jc = [\tau_1, \tau_2]$, on which $\x$ is not AC by construction.
    Then, $\x|_{\bar \Jc}$ is continuous and has Lusin's N-property according to \cite[Theorem 6.12]{gordon1994integrals}, because $\x$ is \ACGs on $\bar \Jc \subset \Ic$.
    Let $M = \sup_{t \in \bar \Jc} \norm{\x(t)}$ be the uniform bound of $\x$ on $\bar\Jc$, and let $Q$ be the uniform bound of $\F$ on the compact subset $\{ \z \in \RR^n : \norm{\z} \le M \} \times \bar\Jc$ of $\Ec$.
Then, the total variation of $\x$ on $\bar\Jc$ is bounded from above by $(\tau_2 - \tau_1) Q$.
    But this implies that $\x$ is AC on $\bar \Jc$, a contradiction.
\end{proof}
\vspace{-\baselineskip}

\endgroup

\bibliographystyle{elsarticle-harv}
\bibliography{literature}           

\begin{thebibliography}{18}
\expandafter\ifx\csname natexlab\endcsname\relax\def\natexlab#1{#1}\fi
\providecommand{\url}[1]{\texttt{#1}}
\providecommand{\href}[2]{#2}
\providecommand{\path}[1]{#1}
\providecommand{\DOIprefix}{doi:}
\providecommand{\ArXivprefix}{arXiv:}
\providecommand{\URLprefix}{URL: }
\providecommand{\Pubmedprefix}{pmid:}
\providecommand{\doi}[1]{\href{http://dx.doi.org/#1}{\path{#1}}}
\providecommand{\Pubmed}[1]{\href{pmid:#1}{\path{#1}}}
\providecommand{\bibinfo}[2]{#2}
\ifx\xfnm\relax \def\xfnm[#1]{\unskip,\space#1}\fi
\bibitem[{Aldana-L{\'o}pez et~al.(2022)Aldana-L{\'o}pez, Seeber,
  G{\'o}mez-Guti{\'e}rrez, Angulo and Defoort}]{aldana2022redesign}
\bibinfo{author}{Aldana-L{\'o}pez, R.}, \bibinfo{author}{Seeber, R.},
  \bibinfo{author}{G{\'o}mez-Guti{\'e}rrez, D.}, \bibinfo{author}{Angulo,
  M.T.}, \bibinfo{author}{Defoort, M.}, \bibinfo{year}{2022}.
\newblock \bibinfo{title}{A redesign methodology generating predefined-time
  differentiators with bounded time-varying gains}.
\newblock \bibinfo{journal}{International Journal of Robust and Nonlinear
  Control} .
\bibitem[{Espitia and Perruquetti(2021)}]{espitia2021predictor}
\bibinfo{author}{Espitia, N.}, \bibinfo{author}{Perruquetti, W.},
  \bibinfo{year}{2021}.
\newblock \bibinfo{title}{Predictor-feedback prescribed-time stabilization of
  {LTI} systems with input delay}.
\newblock \bibinfo{journal}{IEEE Trans. Aut. Control} \bibinfo{volume}{67},
  \bibinfo{pages}{2784--2799}.
\bibitem[{Filippov(1988)}]{filippov1988differential}
\bibinfo{author}{Filippov, A.F.}, \bibinfo{year}{1988}.
\newblock \bibinfo{title}{Differential Equations with Discontinuous Right-Hand
  Side}.
\newblock \bibinfo{publisher}{Kluwer}.
\bibitem[{Gordon(1994)}]{gordon1994integrals}
\bibinfo{author}{Gordon, R.A.}, \bibinfo{year}{1994}.
\newblock \bibinfo{title}{The Integrals of {Lebesgue}, {Denjoy}, {Perron}, and
  {Henstock}}.
\newblock \bibinfo{publisher}{American Math. Society}.
\bibitem[{Holloway and Krstic(2019)}]{holloway2019prescribed}
\bibinfo{author}{Holloway, J.}, \bibinfo{author}{Krstic, M.},
  \bibinfo{year}{2019}.
\newblock \bibinfo{title}{Prescribed-time observers for linear systems in
  observer canonical form}.
\newblock \bibinfo{journal}{IEEE Trans. Aut. Control} \bibinfo{volume}{64},
  \bibinfo{pages}{3905--3912}.
\bibitem[{Kurzweil and Schwabik(1990)}]{kurzweil1990ordinary}
\bibinfo{author}{Kurzweil, J.}, \bibinfo{author}{Schwabik, {\oldv{S}}.},
  \bibinfo{year}{1990}.
\newblock \bibinfo{title}{Ordinary differential equations the solution of which
  are {ACG$_*$}-functions}.
\newblock \bibinfo{journal}{Archivum Mathematicum} \bibinfo{volume}{26},
  \bibinfo{pages}{129--136}.
\bibitem[{Levant(1998)}]{levant1998robust}
\bibinfo{author}{Levant, A.}, \bibinfo{year}{1998}.
\newblock \bibinfo{title}{Robust exact differentiation via sliding mode
  technique}.
\newblock \bibinfo{journal}{Automatica} \bibinfo{volume}{34},
  \bibinfo{pages}{379--384}.
\bibitem[{Levant(2005)}]{levant2005homogeneity}
\bibinfo{author}{Levant, A.}, \bibinfo{year}{2005}.
\newblock \bibinfo{title}{Homogeneity approach to high-order sliding mode
  design}.
\newblock \bibinfo{journal}{Automatica} \bibinfo{volume}{41},
  \bibinfo{pages}{823--830}.
\bibitem[{Levant(2007)}]{levant2007principles}
\bibinfo{author}{Levant, A.}, \bibinfo{year}{2007}.
\newblock \bibinfo{title}{\changed{Principles of 2-sliding mode design}}.
\newblock \bibinfo{journal}{Automatica} \bibinfo{volume}{43},
  \bibinfo{pages}{576--586}.
\bibitem[{Lusin(1912)}]{lusin1912proprietes}
\bibinfo{author}{Lusin, N.}, \bibinfo{year}{1912}.
\newblock \bibinfo{title}{Sur les propri{\'e}t{\'e}s des fonctions mesurables}.
\newblock \bibinfo{journal}{CR Acad. Sci. Paris} \bibinfo{volume}{154},
  \bibinfo{pages}{1688--1690}.
\bibitem[{Polyakov(2011)}]{polyakov2011nonlinear}
\bibinfo{author}{Polyakov, A.}, \bibinfo{year}{2011}.
\newblock \bibinfo{title}{Nonlinear feedback design for fixed-time
  stabilization of linear control systems}.
\newblock \bibinfo{journal}{IEEE Trans. Aut. Control} \bibinfo{volume}{57},
  \bibinfo{pages}{2106--2110}.
\bibitem[{Polyakov and Fridman(2014)}]{polyakov2014stability}
\bibinfo{author}{Polyakov, A.}, \bibinfo{author}{Fridman, L.},
  \bibinfo{year}{2014}.
\newblock \bibinfo{title}{Stability notions and lyapunov functions for sliding
  mode control systems}.
\newblock \bibinfo{journal}{Journal of the Franklin Institute}
  \bibinfo{volume}{351}, \bibinfo{pages}{1831--1865}.
\bibitem[{Roxin(1966)}]{roxin1966finite}
\bibinfo{author}{Roxin, E.}, \bibinfo{year}{1966}.
\newblock \bibinfo{title}{On finite stability in control systems}.
\newblock \bibinfo{journal}{Rend. del Circ. Mat. di Palermo}
  \bibinfo{volume}{15}, \bibinfo{pages}{273--282}.
\bibitem[{Saks(1937)}]{saks1937theory}
\bibinfo{author}{Saks, S.}, \bibinfo{year}{1937}.
\newblock \bibinfo{title}{Theory of the Integral}.
\newblock \bibinfo{publisher}{Hafner Publishing Company}, \bibinfo{address}{New
  York, NY, USA}.
\bibitem[{Schwabik(1990)}]{schwabik1990generalized}
\bibinfo{author}{Schwabik, {\oldv{S}}.}, \bibinfo{year}{1990}.
\newblock \bibinfo{title}{Generalized ordinary differential equations -- a
  survey}, in: \bibinfo{booktitle}{Proc. of the 7th Czechoslovak Conference on
  Differential Equations and Their Applications}, \bibinfo{address}{Prague,
  Czech Republic}. pp. \bibinfo{pages}{59--70}.
\bibitem[{Song et~al.(2017)Song, Wang, Holloway and Krstic}]{song2017time}
\bibinfo{author}{Song, Y.}, \bibinfo{author}{Wang, Y.},
  \bibinfo{author}{Holloway, J.}, \bibinfo{author}{Krstic, M.},
  \bibinfo{year}{2017}.
\newblock \bibinfo{title}{Time-varying feedback for regulation of normal-form
  nonlinear systems in prescribed finite time}.
\newblock \bibinfo{journal}{Automatica} \bibinfo{volume}{83},
  \bibinfo{pages}{243--251}.
\bibitem[{Zhou et~al.(2022)Zhou, Michiels and Chen}]{zhou2022fixed}
\bibinfo{author}{Zhou, B.}, \bibinfo{author}{Michiels, W.},
  \bibinfo{author}{Chen, J.}, \bibinfo{year}{2022}.
\newblock \bibinfo{title}{Fixed-time stabilization of linear delay systems by
  smooth periodic delayed feedback}.
\newblock \bibinfo{journal}{IEEE Transactions on Automatic Control}
  \bibinfo{volume}{67}, \bibinfo{pages}{557--573}.
\bibitem[{Zhou and Shi(2021)}]{zhou2021prescribed}
\bibinfo{author}{Zhou, B.}, \bibinfo{author}{Shi, Y.}, \bibinfo{year}{2021}.
\newblock \bibinfo{title}{Prescribed-time stabilization of a class of nonlinear
  systems by linear time-varying feedback}.
\newblock \bibinfo{journal}{IEEE Trans. Aut. Control} \bibinfo{volume}{66},
  \bibinfo{pages}{6123--6130}.

\end{thebibliography}

\end{document}